\documentclass[12pt]{article}
\usepackage{graphicx}
\usepackage{scicite}
\usepackage{times}

\newenvironment{sciabstract}{%
\begin{quote} \bf}
{\end{quote}}

\newcounter{lastnote}

\title{HD~181068: A Red Giant in a Triply-Eclipsing Compact Hierarchical Triple System}

\author{A. Derekas$^{1,2,3*}$, L.L. Kiss$^{2,4}$, T. Borkovits$^{5,6}$, D. Huber$^{4}$, H. Lehmann$^{7}$,\\
J. Southworth$^{8}$, T.R. Bedding$^{4}$, D. Balam$^{9}$, M. Hartmann$^{6}$, M. Hrudkova$^{6}$, \\
M.J. Ireland$^{4}$, J. Kov\'acs$^{10}$,  Gy. Mez\H{o}$^{2}$, A. Mo\'or$^{2}$, E. Niemczura$^{11}$,\\
G.E. Sarty$^{12}$, Gy.M. Szab\'o$^{2}$, R. Szab\'o$^{2}$, J.H. Telting$^{13}$, A. Tkachenko$^{6}$,\\ 
K. Uytterhoeven$^{14,15}$, J.M. Benk\H o$^{2}$, S.T. Bryson$^{16}$,\\
V. Maestro$^{4}$, A. E. Simon$^{2}$, D. Stello$^{4}$, G. Schaefer$^{17}$, C. Aerts$^{18,19}$,\\ 
T.A. ten Brummelaar$^{17}$,  P. De Cat$^{20}$, H.A. McAlister$^{17}$, C. Maceroni$^{21}$,\\ 
A. M\'erand$^{22}$, M. Still$^{16}$, J. Sturmann$^{17}$, L. Sturmann$^{17}$,\\
 N. Turner$^{17}$,  P.G. Tuthill$^{4}$,  J. Christensen-Dalsgaard$^{23}$, \\ 
R.L. Gilliland$^{24}$, H. Kjeldsen$^{23}$, E.V. Quintana $^{25}$, P. Tenenbaum$^{25}$, J.D. Twicken$^{25}$\\
\\
\scriptsize{$^{1}$Department of Astronomy, E\"otv\"os University, Budapest, Hungary, E-mail: derekas@konkoly.hu}\\
\scriptsize{$^{2}$Konkoly Observatory, Hungarian Acadamey of Sciences, H-1525 Budapest, PO Box 67, Hungary}\\
\scriptsize{$^{3}$Magyary Zolt\'an Postdoctoral Research Fellow}\\
\scriptsize{$^{4}$Sydney Institute for Astronomy (SIfA), School of Physics, University of Sydney, NSW 2006, Australia}\\
\scriptsize{$^{5}$Baja Astronomical Observatory, H-6500 Baja, Szegedi \'ut, Kt. 766, Hungary}\\
\scriptsize{$^{6}$E\"otv\"os J\'ozsef College, H-6500 Baja, Szegedi \'ut 2, Hungary}\\
\scriptsize{$^{7}$Th\"uringer Landessternwarte Tautenburg, Karl-Schwarzschild-Observatorium, 07778 Tautenburg, Germany}\\
\scriptsize{$^{8}$Astrophysics Group, Keele University Newcastle-under-Lyme, ST5 5BG, UK}\\
\scriptsize{$^{9}$Dominion Astrophysical Observatory, Herzberg Institute of Astrophysics,5071 West Saanich Road, Victoria, BC, V9E 2E7, Canada}\\
\scriptsize{$^{10}$Gothard Observatory, E\"otv\"os University, H-9704 Szombathely, Szent Imre Herceg u. 112., Hungary}\\
\scriptsize{$^{11}$Astronomical Institute, Wroc\l aw University, Kopernika 11, 51-622 Wroc\l aw, Poland}\\
\scriptsize{$^{12}$Department of Physics and Engineering Physics, University of Saskatchewan, 9 Campus Drive, Saskatoon, Saskatchewan S7N 5A5, Canada}\\
\scriptsize{$^{13}$Nordic Optical Telescope, Apartado 474, 38700 Santa Cruz de La Palma, Spain}\\
\scriptsize{$^{14}$Lab. AIM, CEA/DSM-CNRS-Universit\'e Paris Diderot; CEA, IRFU, SAp, Saclay, 91191, Gif-sur-Yvette, France}\\
\scriptsize{$^{15}$Kiepenheuer-Institut f\"ur Sonnenphysik, Schöneckstr. 6, 79104 Freiburg, Germany}
\scriptsize{$^{16}$NASA Ames Research Center, Moffett Field, CA 94035, USA}\\
\scriptsize{$^{17}$Center for High Angular Resolution Astronomy, Georgia State University, PO Box 3969, Atlanta, Georgia 30302-3969, USA}\\
\scriptsize{$^{18}$Instituut voor Sterrenkunde, Katholieke Universiteit Leuven, Celestijnenlaan 200 D, 3001 Leuven, Belgium}\\
\scriptsize{$^{19}$IMAPP, Department of Astrophysics, Radboud University Nijmegen, P.O. Box 9010, NL-6500 GL Nijmegen, The Netherlands}\\
\scriptsize{$^{20}$Royal Observatory of Belgium, Ringlaan 3, 1180 Brussel, Belgium}\\
\scriptsize{$^{21}$INAF - Osservatorio astronomico di Roma, via Frascati 33, I-00040 Monteporzio C., Italy}\\
\scriptsize{$^{22}$European Southern Observatory, Alonso de C\'ordova 3107, Casilla 19001, Santiago 19, Chile}\\
\scriptsize{$^{23}$Department of Physics and Astronomy, Building 1520, Aarhus University, 8000 Aarhus C, Denmark}\\
\scriptsize{$^{24}$Space Telescope Science Institute, 3700 San Martin Drive, Baltimore, MD 21218, USA}\\
\scriptsize{$^{25}$SETI Institute, Moffett Field, CA 94035, USA}\\
\\
\scriptsize{$^\ast$To whom correspondence should be addressed; E-mail:  derekas@konkoly.hu}
}

\date{}

\begin{document}

\newcommand{\trinity}{HD~181068}
\newcommand{\sun}{\odot}
\newcommand{\apj}{ApJ}
\newcommand{\mnras}{MNRAS}
\newcommand{\aap}{A\&A}
\newcommand{\aaps}{A\&AS}
\newcommand{\aj}{AJ}

\baselineskip24pt

\maketitle

\begin{sciabstract}

Hierarchical triple systems comprise a close binary and a more distant component. They are important for testing theories of star formation and of stellar evolution in the presence of nearby companions. We obtained 218 days of {\em Kepler} photometry of HD~181068 (magnitude of 7.1), supplemented by ground-based spectroscopy and interferometry, which show it to be a hierarchical triple with two types of mutual eclipses.  The primary is a red giant that is in a 45-day orbit with a pair of red dwarfs in a close 0.9-day orbit.  The red giant shows evidence for tidally-induced oscillations that are driven by the orbital motion of the close pair.  HD~181068 is an ideal target for studies of dynamical evolution and testing tidal friction theories in hierarchical triple systems.

\end{sciabstract}

\noindent The {\em Kepler} space mission is designed to observe continuously more than $10^5$ stars, with the ultimate goal of detecting a sizeable sample of Earth-like planets around main-sequence stars \cite{bor10}. We obtained 218 days of Kepler photometry \cite{koc10,jen10a,jen10b} of HD~181068, a star with magnitude $V=7.1$ and a distance of about 250\,pc.  It has been previously identified as a single-lined spectroscopic binary \cite{gui09} but there have been no reports of eclipses. 

\begin{figure*}
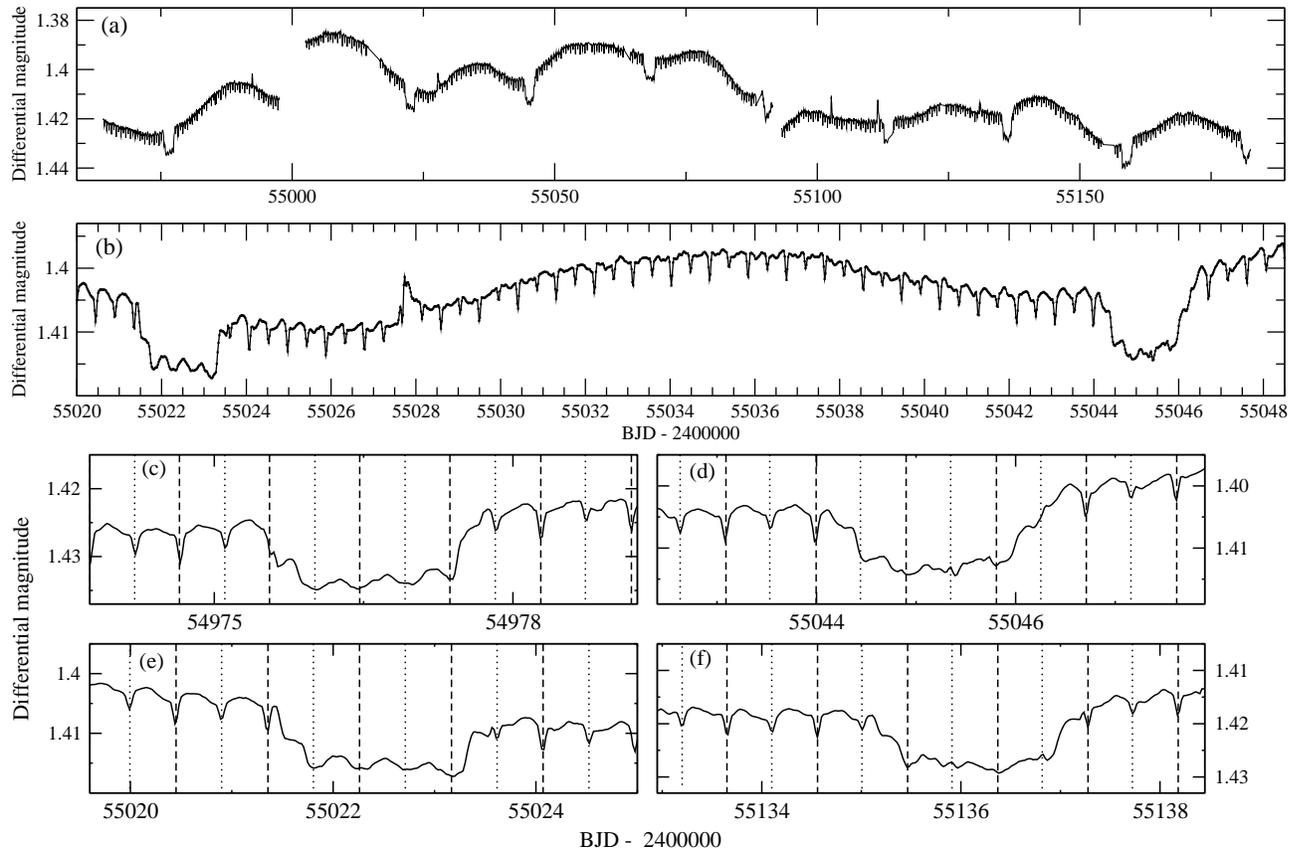

\begin{center}
\includegraphics[width=17cm]{lc5.eps}
{\vskip 1mm}
\includegraphics[width=16.5cm]{minima1.eps}
\caption{\label{lc} {\em Kepler}-band light curve of \trinity\ from observations
in long-cadence mode: (a) the full 218 days; (b)~a 28-d segment showing two
consecutive deep minima. (c--f)~close-ups of two secondary minima and
two primary minima of the 45.5~d eclipses.  
The dashed and dotted lines mark the primary and the secondary minima of the 0.9~d eclipses, respectively. The discontinuities in the top panel correspond to the telescope rolls at the end of each quarter.}
\end{center}
\end{figure*}

The data were obtained in long-cadence (LC) mode (one point every 29.4~minutes) over 218 days using Quarters 1, 2 and 3. Our observations reveal a very distinctive light curve.  It shows eclipses every $\sim$22.7~days and slow variations in the upper envelope (in Fig.~\ref{lc}a) that are likely caused by ellipsoidal distortion of the primary component. There are also very regular and much narrower eclipses (Fig.~\ref{lc}b). These minima have alternating depths, corresponding to a close pair (B and C), with an orbital period of $\sim$0.9~d. The 22.7-d eclipses all have similar depths, but there are subtle differences between consecutive minima. Radial velocity observations (Supporting Online Material, SOM) show that the true orbital period of the BC pair around the A component is 45.5~d. The narrow 0.9-d eclipses essentially disappear during both types of the deep minima, implying that the three stars have very similar surface brightnesses, so that when the BC pair is in front of A, their mutual eclipses do not change the total amount of light coming from the system (in accordance with the nearly-equal depths of the two deep minima). When the BC pair is in front of A, the BC's secondary eclipses appear as tiny brightenings (Fig.~\ref{lc}d and f), showing that the surface brightness of B is almost equal to that of A, while C is a bit fainter, so that its disappearance behind B allows the extra light from A to reach us.

The observed changes in the eclipses of the BC pair and the radial velocity variations of the A component confirm that the A and BC systems are physically associated and not a chance alignment. Their periods are $P_{\rm BC}$=0.90567(2)~d and $P_{\mbox{\scriptsize A-BC}}=45.5178(20)$~d. Given the shallow depths of the eclipses, star A must be the most luminous object in the system.  In addition to the eclipses, there are brightness fluctuations during the long period minima which imply that component A is also an intrinsic variable star with a mean cycle length close to half the shorter orbital period, possibly indicating tidally-induced oscillations.

In addition, there were several flare-like events in the light curve that usually lasted about 6-8 hours. We checked the {\em Kepler} Data Release Notes \cite{note1} for documented instrumental effects in the vicinity of the `flares', but found none. Moreover, almost all flares appear right after the shallower minimum of the BC pair, suggesting that this activity might be related to the close pair.

\begin{table}   
\begin{center}
\caption{\label{tab:orbit}Orbital elements for the wider system derived from the A component's radial velocity curve (Fig.\ \ref{rv}).}
\begin{tabular}{ll}  
\hline\hline
 Element & Value \\ 
\hline 
$P_{\mbox{\scriptsize A-BC}}$ & 45.5178 d (fixed) \\
$T_{\rm Min I}$ &  2455454.573 $\pm$ 0.095 (fixed) \\
$K_{2}$ & 37.195 $\pm$ 0.053 km s$^{-1}$ \\
${\rm v}_{\gamma}$ &  6.993 $\pm$ 0.011  km s$^{-1}$ \\
$e_{2}$     & 0.0 (fixed) \\
$f(m)$ &  0.24 $\pm$ 0.02 M$_{\odot}$ \\
\hline
\end{tabular} 
\end{center}  
\end{table}

\begin{figure}
\begin{center}
\includegraphics[width=8.5cm]{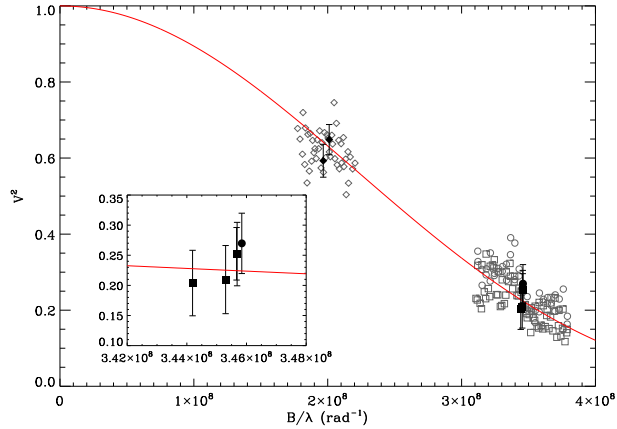}
\caption{\label{pavo} Squared visibility versus spatial frequency from PAVO on CHARA. Grey points show all collected measurements, and black symbols the average of each scan over all wavelength channels. Each symbol type corresponds to a different night of observations. The solid line is the best fitting model. The inset shows a close-up of the observations at the longer baselines.}
\end{center}
\end{figure}

We looked for optically resolved companion(s) with a 1-m telescope (SOM), but found none. We also obtained 41 high-resolution optical spectra to measure the orbital reflex motion of the A component (SOM). The orbital parameters  for the wider system (Table 1) reveal that star A revolves on a circular orbit, which has an orbital period twice the separation of the two consecutive flat-bottomed minima in the light curve (SOM). Long-baseline interferometry using the PAVO beam combiner (Precision
Astronomical Visible Observations, \cite{ire08}) at the CHARA Array (Center for High Angular Resolution Astronomy, \cite{ten05}) show that the angular diameter of \trinity~A, corrected for limb darkening \cite{han74} is  $\theta_{LD}= 0.461\pm 0.011$~milli-arcsecond (Fig.~\ref{pavo}).

Combining the measured angular diameter with the Hipparcos parallax of $4.0\pm0.4$\,mas \cite{vanl07}, we find the linear radius of the primary component to be $R = 12.4 \pm 1.3 R_{\sun}$.  Using the spectroscopically determined $T_{\rm eff}=5100\pm200$\,K, this 
 a luminosity of $L = 93 \pm 19 L_{\sun}$.  This value matches that found from the Hipparcos parallax and the apparent magnitude. We also estimated the absolute magnitude of \trinity~A based on the Wilson-Bappu effect \cite{wil57}, which correlates the width of the chromospheric Ca II K emission line at 3934\,\AA\ with the $V$-band absolute magnitude. Using the latest calibration \cite{pac03}, the measured width of the emission core $W_0$=72.8 km/s implies an absolute brightness of $M_{V}=-0.3$ mag, which matches the parallax and the interferometric results.

We estimated the mass of \trinity~A by comparing the effective temperature and luminosity with evolutionary tracks from the BASTI database \cite{pie04}. We obtained $M_{\rm A}\sim 3.0 \pm 0.4 M_{\odot}$, corresponding to a red giant, possibly in the He-core burning phase of its evolution \cite{gir99}. The full spectral energy distribution, constructed using all published broad-band optical magnitudes and infrared flux values, does not show any excess in comparison to a 5200 K photospheric model, indicating that no detectable circumstellar dust, possibly from mass-loss processes on the red giant branch, is present.

We have constrained the parameters of the BC pair by modelling the short-period eclipses in the Kepler band using the {\sc jktebop} code \cite{sou04,sou05}. The ratio of the radii of the B and C components is poorly constrained at present, partly because of the low sampling rate of the {\it Kepler} long-cadence data. The A component contributes 99.29\% of the system light in the {\em Kepler} passband, and the BC pair contribute 0.44\% and 0.27\%, respectively. Taking the $V$-band absolute magnitude of \trinity\,A to be $M_V{\rm (A)} = -0.3$ and assuming that our results for the {\em Kepler} passband are representative of the $V$-band, we find $M_V{\rm (B)} = 5.6$ and $M_V{\rm (C)} = 6.1$. Such absolute magnitudes indicate spectral types of G8\,V and K1\,V for stars B and C, respectively \cite{cox00}. Since we do not have independent measurements of $T_{\rm eff}$ for the BC pair, we can only estimate their masses based on their spectral types. This indicates that their masses are about 0.7$\pm$0.1~$M_{\odot}$ each (SOM).

One puzzling feature of the system is the short-period fluctuations that have the largest amplitudes when the BC pair is behind star A, while remaining apparent with a slowly changing amplitude in all the other phases of the wide orbit. We have investigated this variability of \trinity\,A with a detailed frequency analysis and a comparison to other red giant stars that have similar properties (SOM). The frequency content of the light curve suggests an intimate link to tidal effects in the triple system, with the first four dominant peaks in the power spectrum identifiable as simple linear combinations of the two orbital frequencies. On the other hand, solar-like oscillations (meaning those excited by near-surface convection, as in the Sun but also observed in red giants) that are expected to produce an equidistant series of peaks in the power spectrum, are not visible, even though all stars with similar parameters in the Kepler database do show clear evidence of these oscillations. In other words, the convectively driven solar-like oscillations that we would expect to see in a giant of this type seem to have been suppressed.

In a recent compilation of 724 triple stars \cite{tok08}, there is only one system with an outer orbital period shorter than that of \trinity\ ($\lambda$~Tau, for which $P_{\rm out}$=33.03 d). \cite{car11} has reported the discovery of KOI-126 with similarly short outer period ($P_{\rm out}$=33.92 d). Extremely compact hierarchical triple systems form a very small
minority of hierarchical triplets, with only 7 of the catalogized 724 systems having outer periods shorter than 150 days. Furthermore, \trinity\  and KOI-126 have the highest outer mass ratios ($\sim$2.1 and 3, respectively, defined as $m_{\rm A}/(m_{\rm B}+m_{\rm C}$) among the known systems. In 97\% of the known hierarchical triplets before the Kepler era, the mass of the close binary exceeded that of the wider companion, and even the larger outer mass ratio remained under 1.5. It is not yet clear if this rarity of such systems is caused by an observational selection effect or has an underlying stellar evolutionary or dynamical explanation.

Its properties make \trinity\ an ideal target for dynamical evolutionary studies, and for testing tidal friction theories. Because of its compactness and its massive primary, we can expect short-term orbital element variations on two different time-scales of 46 days (i.e. with period of $P_{\mbox{\scriptsize A-BC}}$), and approximately 6 years ($P_{\mbox{\scriptsize A-BC}}^2/P_{\rm BC}$), the time-scale of the classical apsidal motion and nodal regression \cite{bro36}, which for the triply eclipsing nature, could be observed relatively easily.

\bibliography{scibib}

\begin{thebibliography}{}
\bibitem[5]{SOM} Materials and methods are available as supporting material on Science Online.
\end{thebibliography}

\bibliographystyle{Science}

\appendix

\newpage

\renewcommand\thefigure{S\arabic{figure}}
\renewcommand\thetable{S\arabic{table}}

\section*{Sections to the Supporting Online Material}

\section{Observations and methods}

\subsection{Lucky Imaging}

We checked \trinity\ for resolved optical companions with lucky imaging on the 1m RCC telescope of the Konkoly Observatory. For this, we took over 100,000 short-exposure frames on 2010 June 28/29 and June 29/30, using an Andor IXon$^{EM}$+888 EMCCD, with exposure times of 30--61\,ms in $UBV(RI)_{\rm C}$ filters. The median seeing was about 1.6$^{\prime\prime}$.  In each
filter we obtained 10,000--30,000 frames, from which the best 0.3\% was selected and combined.  The resulting images show clean single-star profiles with typical FWHM of 0.9$^{\prime\prime}$ in $U$, 0.64$^{\prime\prime}$ in $V$ and 0.45$^{\prime\prime}$ in $I_{\rm C}$. For the latter, we have also determined contrast limits at representative separations. An analysis of simulated artificial stars with a wide range of brightnesses  and separations resulted in the following upper limits to the magnitude difference of a hypothetic optical companion:  0.4$^{\prime\prime}$ -- 2.1 mag; 0.8$^{\prime\prime}$ -- 2.9 mag; 1.2$^{\prime\prime}$ -- 5 mag; 2.0$^{\prime\prime}$ -- 6.4 mag.

\subsection{Spectroscopy}

To measure the orbital motion of \trinity~A, we acquired optical spectra at four different observatories.  We obtained 41 spectra in total, as follows: 6 spectra with the FIES spectrograph at the Nordic Optical Telescope (NOT; resolution 47\,000, wavelength range 3623--7270\,\AA); 14 spectra at the Dominion Astrophysical Observatory (DAO; resolution 10\,000 and wavelength range 4300--4556\,\AA); 16 spectra with the 2-m telescope at the Th\"uringer Landessternwarte (TLS) in Tautenburg (resolution 66\,000, wavelength range 4700--7400\,\AA); and 5 spectra at the McDonald Observatory (McD) using the 2.7m telescope and the Robert G. Tull coud\'e spectrograph (resolution 60\,000 and wavelength range 3700--10000\,\AA).

From the light curve, we know that \trinity\,A contributes almost all the light of the triple system.  We fitted theoretical template spectra from the library of \cite{mun05s} to a NOT spectrum and determined the following parameters: $T_{\rm
eff}=5100\pm200$\,K, $\log{g}=2.8\pm0.3$, $[M/H]=-0.6\pm0.3$ and $v~\sin i=14$\,km~s$^{-1}$.  These values are all in good agreement with those of \cite{gui09s}.  As a check, we can also use Str\"omgren photometry from \cite{ols93s}: $V = 7.091$, $b-y = 0.586$, $m_{1} = 0.296$ and $c_{1} = 0.399$.  From these quantities and the calibration of \cite{moo85s}, we find: $T_{\rm eff} = 5200$\,K, $E(b-y) = 0.087$ and $(b-y)_0 = 0.499$.  These numbers are consistent with the spectroscopic results and confirm that \trinity\,A is a G-type giant star.

We have used one high-quality NOT spectrum to measure the width of the Ca II K emission line at 3934 \AA. This has been done interactively with the IRAF task {\sc splot} after converting the wavelength scale to Doppler velocities around the core of the emission. The measured value is $W_{0}=72.8$ km~s$^{-1}$, which was then used to calculate the $V$-band absolute magnitude via the calibration of the Wilson-Bappu effect by \cite{pac03}.

Radial velocities were determined using the IRAF task {\sc fxcor}. The template for all but the TLS spectra was selected from \cite{mun05s}, with closely matching parameters, which ensured that no systematic errors were introduced by spectral template mismatch.  The 16 TLS RVs have been determined in a first step from cross-correlation with the mean spectrum that was iteratively built from the single spectra by shift-and-add according to the measured RVs. This mean spectrum was then analysed using the program {\sc LLmodels} \cite{shu04s} to compute a grid of stellar atmosphere models and the program {\sc SynthV} \cite{tsy96s} to compute the synthetic spectra. We found $T_{\rm eff}=5300\pm100 K$, $\log{g}=2.8\pm0.2$ dex, $[M/H]=-0.2\pm0.1$, and $v~\sin i=14\pm1$ km~s$^{-1}$. Finally, we used the best-fit synthetic spectrum as a template for cross-correlation to
determine the RVs of the TLS spectra on an absolute scale.  Depending on the instrument and the spectra, the velocities are accurate to $\pm$0.5--2 km~s$^{-1}$. The observed radial velocities and their deviations from the orbital fit are listed in Table\ \ref{rvlist}.

\begin{figure}
\begin{center}
\includegraphics[width=8.5cm]{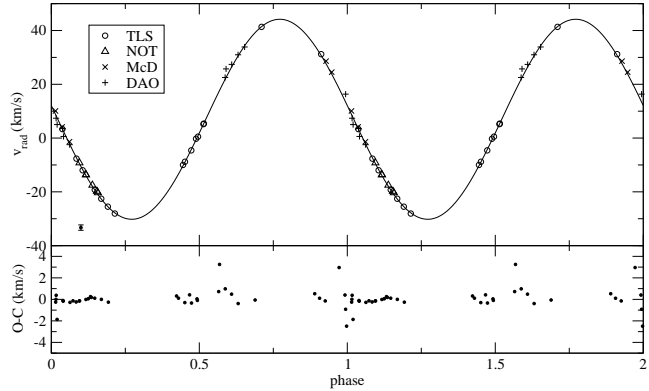}
\caption{\label{rv} Measured RVs versus the orbital phase (45.5~d) for the TLS (circles), NOT (triangles), McDonald (McD, crosses), and DAO (pluses) observations. The vertical bar in the lower left corner shows the size of the representative
$\pm$1 km~s$^{-1}$ uncertainty.}
\end{center}
\end{figure}

\begin{table} \begin{center}
\caption{\label{rvlist} The list of radial velocity measurements. The typical uncertainty is $\pm$1 km~s$^{-1}$. Sources: 1=TLS, 2=NOT, 3=MCD, 4=DAO}
\begin{tabular}{lrrrl} \hline \hline
BJD 2455000+     &  phase &    RV (km~s$^{-1}$) &    O$-$C & Source \\
\hline
358.53084 & 0.8900 & 31.19 & 0.52 & 1 \\
369.55796 & 0.1323 & $-$20.24 & 0.26 & 2 \\
369.69747 & 0.1353 & $-$20.79 & 0.18 & 2 \\
404.79220 & 0.9063 & 27.72 & 0.11 & 3 \\
405.66236 & 0.9255 & 23.61 & $-$0.15 & 3 \\
407.80097 & 0.9724 & 16.33 & 2.96 & 4 \\
408.70024 & 0.9922 & 9.19 & 0.40 & 3 \\
408.79660 & 0.9943 & 7.37 & $-$0.92 & 4 \\
408.95076 & 0.9977 & 5.02 & $-$2.49 & 4 \\
409.73019 & 0.0148 & 3.26 & $-$0.26 & 3 \\
409.79654 & 0.0163 & 3.55 & 0.38 & 4 \\
409.93630 & 0.0194 & 0.59 & $-$1.87 & 4 \\
410.85279 & 0.0395 & $-$2.27 & $-$0.10 & 3 \\
410.87936 & 0.0401 & $-$2.46 & $-$0.17 & 4 \\
412.38515 & 0.0732 & $-$9.66 & $-$0.14 & 2 \\
413.37500 & 0.0949 & $-$14.06 & $-$0.14 & 2 \\
413.39495 & 0.0953 & $-$14.14 & $-$0.14 & 2 \\
414.37817 & 0.1169 & $-$17.97 & $-$0.01 & 2 \\
428.30911 & 0.4230 & $-$10.01 & 0.32 & 1 \\
428.61223 & 0.4297 & $-$8.82 & 0.11 & 1 \\
429.58763 & 0.4511 & $-$4.58 & $-$0.30 & 1 \\
430.31478 & 0.4671 & $-$0.25 & 0.42 & 1 \\
430.61761 & 0.4737 & 0.52 & $-$0.34 & 1 \\
431.45537 & 0.4921 & 5.18 & 0.05 & 1 \\
431.52041 & 0.4935 & 5.37 & $-$0.09 & 1 \\
434.78970 & 0.5654 & 22.54 & 0.73 & 4 \\
434.92856 & 0.5684 & 25.72 & 3.25 & 4 \\
435.80118 & 0.5876 & 27.40 & 0.98 & 4 \\
436.77665 & 0.6090 & 30.98 & 0.48 & 4 \\
437.78550 & 0.6312 & 33.88 & $-$0.39 & 4 \\
440.39439 & 0.6885 & 41.35 & $-$0.06 & 1 \\
455.28666 & 0.0157 & 3.32 & 0.01 & 1 \\
457.44107 & 0.0630 & $-$7.65 & $-$0.28 & 1 \\
458.39301 & 0.0839 & $-$11.98 & $-$0.24 & 1 \\
460.25545 & 0.1248 & $-$19.24 & 0.06 & 1 \\
461.25709 & 0.1468 & $-$22.56 & 0.11 & 1 \\
462.27140 & 0.1691 & $-$25.52 & 0.00 & 1 \\
463.34313 & 0.1927 & $-$28.09 & $-$0.26 & 1 \\
\hline \end{tabular} \end{center} \end{table}

The orbital solution (see Fig.~\ref{rv}) was calculated by the method of differential corrections.  We omitted the DAO RVs because they show a much larger scatter around the calculated orbital curve, despite being observed during the same epoch as the other instruments. The eccentricity from the fit was $e=0.022\pm0.023$, which is consistent with zero, and in the final solution we set $e=0$ because its inclusion as a free parameter did not improve the solution. We also fixed the orbital period to that obtained from the light-curve fitting (a free search gives a slightly different value but does not improve the quality of the solution). Table~\ref{tab:orbit} lists the derived orbital elements. In the corresponding solution, we corrected the NOT RVs by $-$0.35 km~s$^{-1}$ and the McDonald RVs by $-$0.70 km~s$^{-1}$ with respect to the TLS RVs.  This correction minimized the rms to 214 m~s$^{-1}$, and its influence on the derived elements was only marginal.

\subsection{Interferometry}

Interferometric observations were performed on three nights in 2010 July, using two different baselines (156.3~m and 248.1~m) of the CHARA Array. All the observations were performed outside the long-period eclipses, meaning that some flux from the BC pair was present during all observations. With an eclipse depth of only 1\%, however, the companions are much fainter than the primary and are negligible in our analysis. The raw data were reduced using the PAVO data analysis pipeline. 

We obtained 7 scans of HD181068 over 23 wavelength channels spanning from 0.65 to 0.8 mum each, yielding a total of 161 V2 measurements. We observed six stars of spectral type A0 to calibrate the raw visibilities. All stars were located within $5^{\circ}$ in the sky and their estimated diameters were at least a factor of 2 smaller than HD 181068A. Inspection of the data revealed that three of the calibrators (HD179395, HD182487 and HD181521) are potential binaries and could not be used for calibration. The remaining three stars used to calibrate our data were HD179733, HD180138 and HD184787. To determine the angular diameter of HD181068 A a limb-darkened disc model \cite{hanb74} was fitted to the calibrated visibilities. The corresponding linear limb-darkening coefficient was determined by interpolating the spectroscopically determined values of $\log~g$ and $T_{\rm eff}$ in the grid of \cite{claret00}, yielding $\mu = 0.63 \pm 0.02$. The resulted angular diameter of HD181068 A is $\theta_{LD}=0.461\pm 0.011$ milli-arcsecond.

The uncertainty on the diameter was estimated using 40000 Monte-Carlo simulations as follows: For each simulation, we drew realizations corresponding to Gaussian distributions for the calibrator angular diameters, limb-darkening coefficient and wavelength channels, with assumed standard deviations of 5\%, 3\% and 0.5\%. With these parameters we then calibrated the raw visibility measurements and fitted an angular diameter to the calibrated data using least-squares minimization. To account for random errors, including correlations between wavelength channels, we perturbed this fit by adding random numbers generated from the empirical covariance matrix of the data and then repeated the fit \cite{pre92}. The final uncertainty was taken as the standard deviation of the resulting total distribution, scaled by square root of the reduced $\chi^2$ value (2.5) as determined from the fit to the original data. The measurements of the 7 scans averaged over 23 wavelength channels are listed in Table\ \ref{inter}.

\begin{table} \begin{center}
\caption{\label{inter} Calibrated interferometric measurements of HD181068 averaged over 23 wavelength channels for each scan. The full list of measurements are available from the authors on request.}
\begin{tabular}{ccccc} \hline \hline
JD-2451545 & Projected Baseline (m) & Spatial Frequency ($\times 10^8 rad^{-1}$) & avg(V$^2$) & avg($\sigma V^2$) \\
\hline
3837.973 &  144.064 & 2.010901  &   0.648  &   0.040\\
3838.000 &  140.892 & 1.966629  &   0.593  &   0.043\\
3852.920 &  247.766 & 3.458418  &   0.270  &   0.050\\
3852.930 &  247.654 & 3.456854  &   0.252  &   0.053\\
3853.929 &  247.640 & 3.456666  &   0.252  &   0.044\\
3853.944 &  247.378 & 3.452999  &   0.210  &   0.057\\
3853.969 &  246.584 & 3.441913  &   0.204  &   0.054\\
\hline \end{tabular} \end{center} \end{table}

\subsection{\trinity\ B and C}

We have constrained the parameters of the BC pair by modelling the short-period eclipses in the Kepler band. First, we removed the long-term variations of the uneclipsed brightness in the light curve by fitting spline function polynomials and removing data obtained during the long-period eclipses. In this way, the light from star A was assigned to be the `third light' component. The parameters of the BC pair were deduced by modelling the short-period eclipses in the Kepler band using the {\sc jktebop} code \cite{sou04a,sou05a}.
A preliminary fit was performed, allowing a few outlying data points to be identified and removed. A detailed fit was then made, using numerical integration to account for the 30-minute long duration of individual observations. Uncertainties in the parameters of the fit were calculated using 1000 Monte Carlo simulations \cite{sou05s}. In each simulation, a synthetic dataset was created by evaluating the best-fitting model at the observed times and adding Gaussian noise. This was then fitted in the same way as the real data, starting from initial parameter values which were perturbed versions of the best-fitting parameter values. The error estimate for each fitted parameter was then evaluated by taking the inner 68.3\% of all the values of the parameter found from the synthetic datasets. The resulting photometric parameters are given in Table~\ref{tab:lc}.

\begin{table} \begin{center}
\caption{\label{tab:lc} Photometric parameters obtained for the short-period eclipses, using standard symbols in the study of eclipsing binary systems. $L_{\rm A}$,$L_{\rm B}$ and $L_{\rm C}$ are all expressed as fractions of the total light of the system in the Kepler passband.}
\begin{tabular}{lll} \hline \hline
Parameter & Best fit & Uncertainty \\
\hline
$P_{\rm BC}$ (d)                        & 0.9056770 & 0.0000026 \\
$T_{\rm Min I}$ (BJD)               &2455051.23625& 0.00020 \\
$i_{1}$ (degrees)              & 87.7      & 1.6       \\
$(R_{\rm B}+R_{\rm C})/a_{\rm BC}$  & 0.3288    & 0.0044    \\
$R_{\rm C}/R_{\rm B}$               & 1.01      & 0.13      \\
$R_{\rm B}/a_{\rm BC}$              & 0.164     & 0.011     \\
$R_{\rm C}/a_{\rm BC}$              & 0.165     & 0.011     \\
$L_{\rm A}$                         & 0.9929    & 0.0006    \\
$L_{\rm B}$                         & 0.0044    & 0.0007    \\
$L_{\rm C}$                         & 0.0027    & 0.0004    \\

\hline \end{tabular} \end{center} \end{table}

We found the ratio of the radii of the B and C components to be poorly constrained at present, partly due to the low sampling rate of the {\it Kepler} long-cadence data. The A component contributes 99.29\% of the system light in the {\em Kepler} passband, and the BC pair contribute 0.44\% and 0.27\%, respectively. Taking the $V$-band absolute magnitude of \trinity\,A to be $M_V{\rm (A)} = -0.3$ and assuming that our results for the {\em Kepler} passband are representative of the $V$-band, we find $M_V{\rm (B)} = 5.6$ and $M_V{\rm (C)} = 6.1$. Such absolute magnitudes indicate spectral types of G8\,V and K1\,V for stars B and C, respectively.

\begin{figure}
\begin{center}
\includegraphics[width=8.5cm]{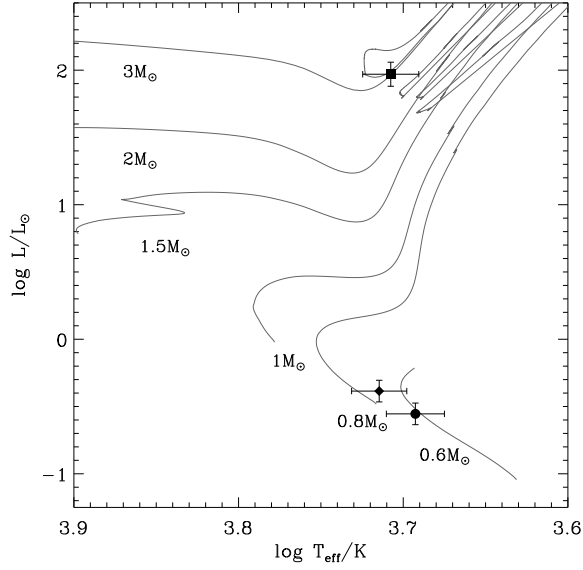}
\caption{\label{hrd} A Hertzsprung-Russell Diagram using the BASTI tracks \cite{piet04} for $[M/H]=-0.25$. It shows the locations of components A, B and C, marked by the filled square, diamond and circle, respectively. }
\end{center}
\end{figure}

We can only estimate masses of the BC pair based on the spectral type because there is no independent $T_{\rm eff}$ measurement of them. Fig.\ \ref{hrd} shows a Hertzsprung-Russel Diagram with BASTI evolutionary tracks \cite{piet04} of different masses for $[M/H]=-0.25$ (as suggested from the spectroscopy), as well as the most likely locations of the components of the triple system. The position of the BC components was derived based on the spectral types stated above, assuming an error of 200~K in $T_{\rm eff}$ and 0.2~mag in absolute magnitude, respectively.

\section{The variability of \trinity~A}

The light curve (see Fig.~\ref{lc}) shows slow variations with the same timescale as the long-period eclipses, which presumably arise from ellipsoidal distortion of the primary.  We also see faster oscillations with the same timescale as the orbital period of the BC pair. These are visible both outside and during the eclipses and are less obvious to interpret.

\begin{figure}
\begin{center}
\includegraphics[width=14cm]{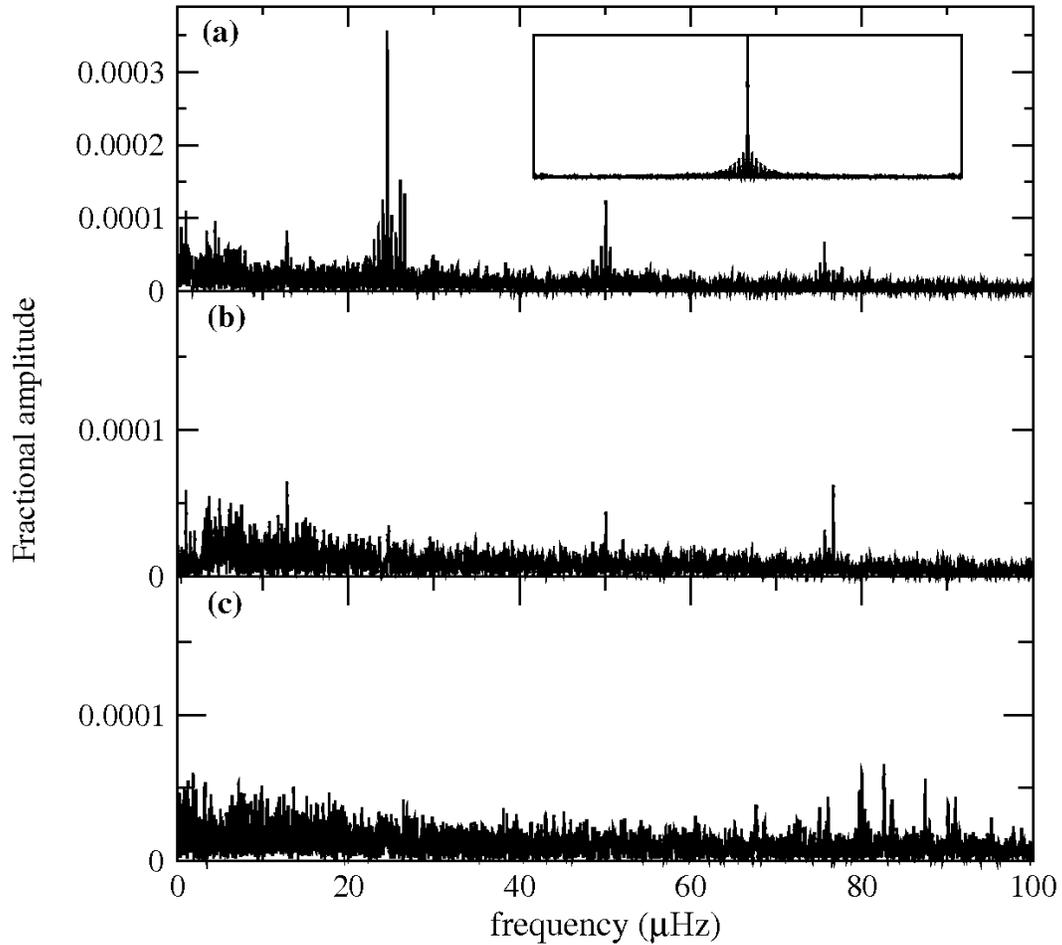}
\caption{\label{fourier} The amplitude spectrum of \trinity~A and the change after 
five prewhitening steps (panels a and b). The inset shows the spectral window function with the same axis scales. Panel c: the spectrum of a typical red giant star (KIC~12507577), for which $\nu_{\rm max}\sim80~\mu$Hz. The vertical scale in panels b and c is increased by
a factor of two.}
\label{distr}
\end{center}
\end{figure}

To investigate this further, Fig.~\ref{fourier}{\em a} shows the amplitude spectrum of the light curve after first removing observations made during both the long- and short-period eclipses.  This procedure left a light curve with a duty cycle slightly above 60\%, and it introduced alias peaks in the amplitude spectrum at the multiples of the orbital frequencies.  The strongest peak in the spectrum occurs at $25~\mu$Hz, corresponding to half the orbital period of the BC binary. As mentioned, this periodicity is clearly visible in the light curve.

To search for other frequencies we used iterative sine-wave fitting (prewhitening) with Period04 \cite{len05s}. In five steps, we measured and identified the following frequencies:
\begin{itemize}
\parskip=-3pt
\item $f_{1}=24.54\,\mu{\rm Hz} = 2(f_{\rm short}-2f_{\rm long})$

\item $f_{2}=25.05\,\mu{\rm Hz} = 2(f_{\rm short}-f_{\rm long})$

\item  $f_{3}=25.56\,\mu{\rm Hz} = 2f_{\rm short}$

\item $f_{4}=51.12\,\mu{\rm Hz} = 4f_{\rm short}$

\item $f_{5}=12.83\,\mu{\rm Hz} = f_{\rm short}+1/T_{\rm obs}$

\end{itemize}

Here, $f_{\rm short}=1/P_{\rm BC}$, $f_{\rm long}=1/P_{\mbox{\scriptsize A-BC}}$, and $T_{\rm obs}$ is the time span of observations. Fig.~\ref{fourier}{\em b} shows the amplitude spectrum after the slow variations and these five strongest peaks have been subtracted from the time series.

Could the observed signal in \trinity\ arise from solar-like oscillations? According to the parameters derived from the interferometry and spectroscopy, \trinity~A is a red giant star located in the bottom of the red giant branch.  Studies of similar red giants with {\em Kepler} \cite{hub10s,kal10s} show that essentially all stars in this region of the H-R diagram exhibit solar-like oscillations that appear as a broad power excess centred at a characteristic frequency $\nu_{\rm max}$ and composed of a regularly spaced series of peaks. Using the scaling relation of \cite{kje95s} with the derived values of mass, radius and effective temperature, we  estimate $\nu_{\rm max}=64\pm16~\mu$Hz for \trinity~A, while the expected amplitude is about 80 ppm.

For comparison, Fig.~\ref{fourier}{\em c} shows the amplitude spectrum of a typical red giant with $\nu_{\rm max}\sim80~\mu$Hz (KIC~12507577).  To make the comparison exact, we calculated this amplitude spectrum using exactly the same portions of the
light curve that were used for \trinity\ in Fig.~\ref{fourier}{\em a}. In KIC~12507577 we see the regular peaks that characterize solar-like oscillations, whereas in \trinity~A we see just a few peaks whose removal (Fig.~\ref{fourier}{\em b}) leaves only a slowly rising power distribution. The observed signal in \trinity~A is clearly not compatible with solar-like
oscillations.  Indeed, the solar-like oscillations that we would expect to see in a giant of this type seem to have been suppressed.

The frequency content of the light curve suggests an intimate link to the orbital frequencies in the triple system.  We are led to suggest that we are seeing tidally-induced oscillations that are driven by the orbital motion of the BC pair.  Tidally-induced oscillations have previously been reported in a few binary systems \cite{decat00s,han02s,mac09s}, but here the situation is different because the period of the oscillations does not correspond to the orbit of the A component, but rather to that of the
BC pair.  While a fuller discussion of this possibility is postponed to a future publication, we note that the amplitude of tidally driven oscillations can be simply estimated by assuming that the brightness changes are proportional to the tidal heights given by $R_{A}\left( (M_{B}+M_{C})/M_{A}\right) (R_{A}/a)^{3}$, where $a$ is the semimajor axis of the outer binary. Every number put together yield an estimated amplitude of 3 ppt, which is a factor of 10 larger than what we observe and a factor of 40 larger than the amplitude of solar-like oscillations.

\section{Supplemented data files}

In addition to this document describing the supporting extra material, we also make all the data mentioned throughout the paper and SOM available to the general community at the webpage of the journal. The full list of datafiles that are stored in a single compressed tarball file is the following:

\bigskip

{\tt \footnotesize
\begin{tabular}{ll}
1201762som\_sed.txt  & Spectral Energy Distribution (Main text)\\
1201762som\_lucky\_image.fits & I-band lucky image (SOM, 1.1)\\
1201762som\_fies(1..6).fits & 6 NOT/FIES spectra (SOM, 1.2)\\
1201762som\_dao(1..11).fits & 11 DAO spectra (SOM, 1.2)\\
1201762som\_tls(1..16).fits & 16 TLS spectra (SOM, 1.2)\\
                                                  & (barycentric corrections applied)\\
1201762som\_mcd(1..5).fits & 5 McDonald spectra (SOM, 1.2)\\
\end{tabular}
}

\end{document}